\documentclass[aps,pre,reprint,superscriptaddress,fleqn]{revtex4-2}

\usepackage{amsmath, bm, mathtools, color}
\usepackage{graphicx}
\usepackage{CJKutf8}   
\usepackage{hyperref}
\usepackage{enumitem}
\usepackage{multirow}

\usepackage[normalem]{ulem}
\usepackage{xcolor}

\allowdisplaybreaks
\begin{document}
\suppressfloats[t]


\title{Early warning signals for synchronization transitions from partial observations}

\author{Yusuke Kato}
\email{yusukeka@umich.edu}
\affiliation{Gilbert S.~Omenn Department of Computational Medicine and Bioinformatics, University of Michigan, Ann Arbor, 48109-2218, MI, USA}

\author{Naoki Masuda}
\email{naokimas@umich.edu}
\affiliation{Gilbert S.~Omenn Department of Computational Medicine and Bioinformatics, University of Michigan, Ann Arbor, 48109-2218, MI, USA}
\affiliation{Department of Mathematics, University of Michigan, Ann Arbor, 48109-1043, MI, USA}
\affiliation{Center for Computational Social Science, Kobe University, Kobe, 657-8501,
Hyogo, Japan}



\date{\today}

\begin{abstract}
Anticipating the onset of collective synchronization is important in many networked systems, yet observing every oscillator is often impractical. We investigate whether synchronization transitions can be detected from a small set of monitored, or sentinel, nodes. Using a stochastic Kuramoto model on networks, we numerically compare three early warning signals: the local order parameter, its temporal variance, and the variance of individual oscillator phases after removing their mean rotational trends. We also compare sentinel-selection strategies based on node dynamics, degree, and random sampling. We show that, under partial observation, the local order parameter and the variance of detrended phases provide substantially stronger warning signals than the variance of the local order parameter. Selecting nodes according to their dynamical behavior near the transition consistently improves performance over other sentinel-selection methods. With only $\lfloor\ln N\rfloor$ dynamically selected sentinels, two success indicators approach the performance obtained by observing all $N$ nodes. These results demonstrate that synchronization transitions can be anticipated from sparse observations when the warning signal and monitored nodes are chosen appropriately.
\end{abstract}


\maketitle

\section{Introduction}

Various real-world systems undergo abrupt qualitative changes as external environments gradually vary~\cite{feudel_multistability_2018,ratajczak_abrupt_2018,dakos_tipping_2024}. Examples include
ecosystem collapse and species extinction~\cite{scheffer2009early}, rapid shifts
in the climate system~\cite{dakos2008slowing,boers2021observation}, large-scale
failures of power grids~\cite{simpson-porco_voltage_2016}, and the onset of
psychological disorders~\cite{van2014critical,wichers_critical_2016}. Measurable changes that precede
such abrupt transitions, if any, can serve as early warning signals (EWSs) and may
provide time for intervention. A widely studied mechanism behind EWSs is
critical slowing down~\cite{scheffer2009early, scheffer_anticipating_2012,kuehn_mathematical_2011}. Under critical slowing down, a system recovers
increasingly slowly from small perturbations as it approaches a loss of
resilience that corresponds to a bifurcation of a dynamical system underlying the observed data. This loss of resilience can produce characteristic statistical
signatures in the data, including increases in variance and lagged autocorrelation of the observed time series, that
can serve as EWSs for impending transitions~\cite{scheffer2009early, scheffer_anticipating_2012}.

Many complex systems for which anticipating critical transitions is desired, such as ecosystems and power grids, consist of large numbers of interacting
components connected as networks. Therefore, a growing body of research has
extended early-warning approaches to networked dynamical
systems~\cite{dakos_critical_2014,jiang_predicting_2018,aparicio_structure-based_2021,maclaren2023early, pirani2023network,masuda2024anticipating,maclaren2025applicability}. A key insight when considering EWSs for dynamics on networks is that
different nodes carry different information due to the heterogeneous structure of almost all empirical networks.
Moreover, observing the states
of all nodes is costly and often infeasible in practical applications. These considerations
have motivated recent studies on selecting appropriate sentinel nodes, a small set of monitored nodes, for
computing EWSs~\cite{aparicio_structure-based_2021,maclaren2023early,masuda2024anticipating, Yu2026Usingcovariance,maclaren2025applicability}.

A collective phenomenon for which EWSs have been relatively scarcely examined is synchronization of self-oscillatory elements, which is commonly observed in nature and engineering systems in particular~\cite{pikovsky_synchronization_2001}. Transitions between synchronized and desynchronized regimes can have practical system-wide consequences, including failures and functional disruptions. For example, loss of frequency synchrony in power
grids can contribute to cascading failures and large-scale
outages~\cite{witthaut2022collective};
reduced synchrony among clock cells in the central nervous system can weaken
circadian rhythms~\cite{aton_vasoactive_2005}; and collective synchronization
among pedestrians can induce large lateral oscillations in
bridges~\cite{strogatz_crowd_2005}. These potential consequences motivate the development of methods for anticipating synchronization transitions~\cite{fan_anticipating_2021}.

Several studies have sought to anticipate synchronization transitions in coupled-oscillator systems~\cite{garcia2017enhancement,rahjerdi2022indicating, liu2024early, pirani2023network,fan_anticipating_2021, ghosh2022early, leyva_inferring_2020, ghosh_anticipating_2022,leyva2025local}. These studies have developed EWSs based on temporal statistics of macroscopic order parameters such as their variance and skewness~\cite{garcia2017enhancement,rahjerdi2022indicating}, pairwise lag times between oscillators~\cite{leyva_inferring_2020}, the eigenvector associated with the smallest nonzero eigenvalue of the Laplacian matrix~\cite{pirani2023network}, measures derived from information theory~\cite{ghosh2022early,ghosh_anticipating_2022,leyva2025local}, and machine learning methods~\cite{liu2024early,fan_anticipating_2021}.
In fact, most of the previous EWSs for synchronization transitions were computed using observations from all nodes~\cite{garcia2017enhancement,rahjerdi2022indicating,pirani2023network, fan_anticipating_2021,liu2024early, ghosh2022early, ghosh_anticipating_2022, leyva_inferring_2020}; detection of synchronization transitions from partial node observations has scarcely been explored. Leyva et al.\,proposed an EWS computable from partial observations of nodes and compared its performance between high- and low-degree observed nodes~\cite{leyva2025local}. To induce explosive synchronization, they set each node's natural frequency, which characterizes its intrinsic dynamics, primarily according to the node's degree. It therefore remains unknown how synchronization transitions can be anticipated from partial observations when intrinsic node dynamics are not particularly 
assumed to be associated with node degree. It is also generally unknown how one can select informative sentinel nodes for computing EWSs for synchronization transitions.

Motivated by these research gaps, in this study, we develop EWS methods applicable to dynamics on networks for anticipating the onset of synchronization transitions, assuming measurement at a small number of sentinel nodes.
We focus on coupled Kuramoto oscillators on networks, a canonical model for synchronization~\cite{kuramoto_chemical_1984, arenas2008synchronization, rodrigues_kuramoto_2016}. We compare two existing EWSs and propose a third EWS. The proposed EWS captures fluctuations around each oscillator's average rotation and is a natural extension of a popular EWS for non-oscillatory stochastic dynamics. We numerically show that the proposed EWS improves over the existing methods in terms of transition detection rates and that its performance in the case of logarithmically small numbers of nodes
can be close to that in the case of all-node observation.

\section{Methods}
\label{sec:methods}

\begin{figure*}
\centering
\includegraphics[width=.9\linewidth]{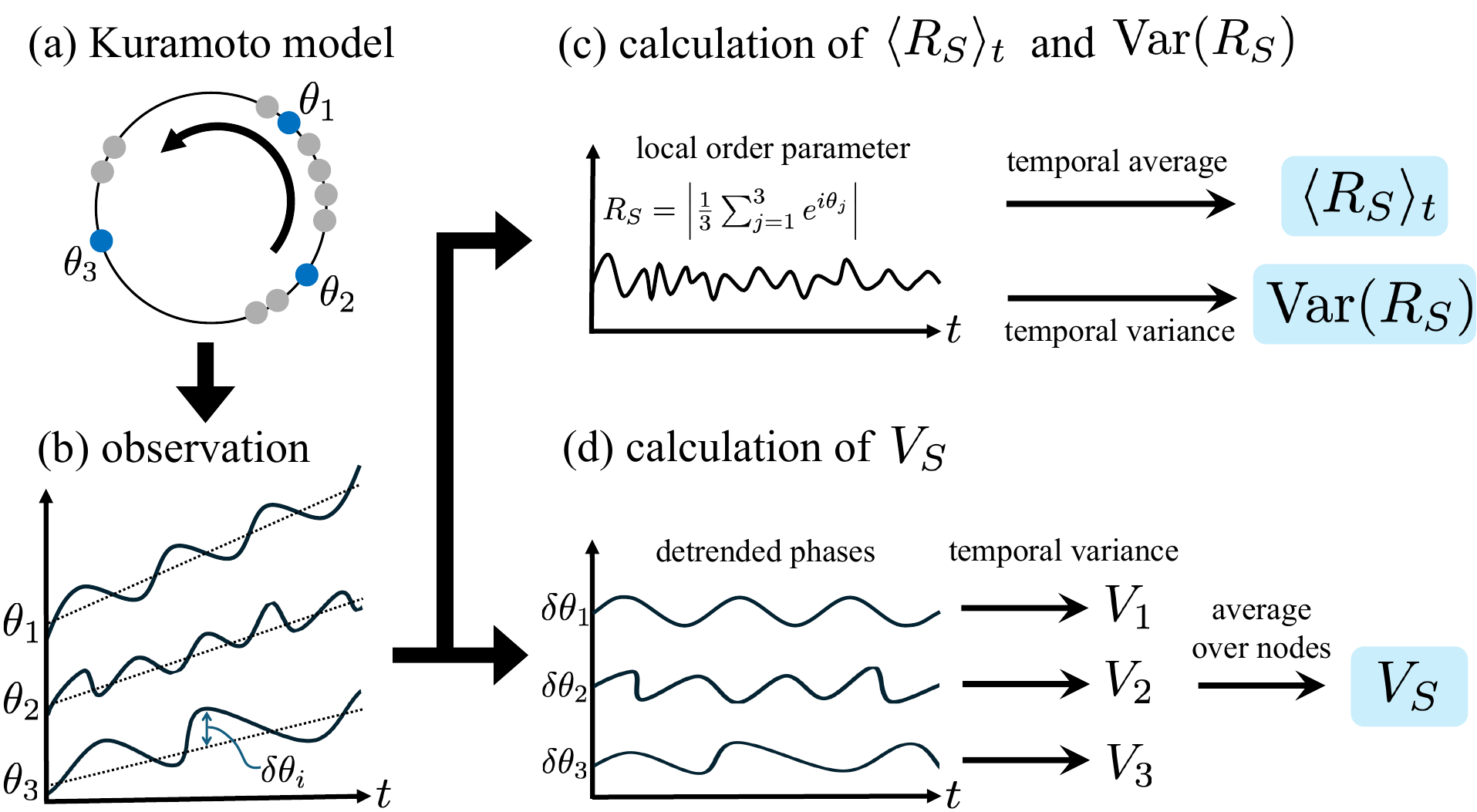}
\caption{Schematic illustration of the calculation of the three EWSs, $\langle R_S \rangle_t, \, \mathrm{Var}(R_S)$, and $V_S$. Panel (a) represents the phase variables $\theta_i$ of the Kuramoto
model [Eq.~\eqref{eq:kura_network}] as filled circles rotating on a unit circle. The three
blue circles represent the observed nodes, i.e., the sentinel nodes. In (b), the solid curves
show typical time courses of the phase variables, $\theta_i$, for the sentinel nodes. The dotted lines represent
their fitted linear trends. The difference between each solid curve and
the corresponding dotted line gives the detrended phase $\delta\theta_i$
in Eq.~\eqref{eq:micro_pre}. Panel (c) shows a typical time course of the local
order parameter $R_S$. We calculate the two EWSs, $\langle R_S\rangle_t$ and
$\mathrm{Var}(R_S)$, by taking the temporal average and variance of $R_S$,
respectively. Panel (d) shows typical time courses of $\delta\theta_i$ for the
sentinel nodes. We calculate the third EWS, $V_S$, as the temporal
variance of each detrended phase, which is averaged over the sentinel nodes.}
\label{fig:scheme}
\end{figure*}

\subsection{Model}
We consider the Kuramoto model with dynamical noise on a network given by
\begin{equation}
    \label{eq:kura_network}
    \dot \theta_i =
        \omega_i
        +
        \frac{K}{\langle k\rangle}
        \sum_{j=1}^{N}
        A_{ij}\sin(\theta_j-\theta_i)
    + \sigma \xi_i(t).
\end{equation}
See
Figs.~\ref{fig:scheme}(a) and (b) for schematics.
Here, $\theta_i$ and $\omega_i$ denote the phase and natural frequency of
oscillator $i$, respectively, $N$ is the number of oscillators (nodes), $K$ is
the coupling strength, $A_{ij}$ is the adjacency matrix,
$\langle k\rangle \coloneqq N^{-1}\sum_{i=1}^N \sum_{j=1}^N A_{ij}$ is the mean degree of the
network, and $\sigma$ is the noise strength. The dot denotes differentiation
with respect to time (i.e., $\dot\theta_i\coloneqq d\theta_i/dt$). We draw each $\omega_i$ independently from a common probability distribution $g(\omega)$. The
noise terms $\xi_i(t)$ are independent Gaussian white-noise processes satisfying
$\langle\xi_i(t)\xi_j(t')\rangle=\delta_{ij}\delta(t-t')$.
The division by $\langle k\rangle$ places all
networks on a comparable coupling scale. Although Eq.~\eqref{eq:kura_network} is well defined for directed and weighted networks, we restrict ourselves to undirected unweighted networks (i.e., $A_{ij}=A_{ji}\in\{0,1\}$) for simplicity.

\subsection{Networks}
\label{sec:networks}

We use three model networks with $N=100$ and six empirical networks. The model networks are the complete graph (all-to-all coupling), an Erd\H{o}s--R\'enyi (ER) random graph~\cite{erdos1959random}, and
a scale-free network. We regenerate the ER and scale-free networks until they are connected. We use one instance of each model network
throughout the study. The empirical networks are arbitrarily selected from the
\texttt{networkdata} R package~\cite{schoch2022networkdata}, and we reduce each network to its
largest connected component. 

For each network, Table~\ref{tab:kc} lists the number of nodes $N$, number of
edges $|E|$, and mean degree $\langle k \rangle$. We provide further details of the networks
in Appendix~\ref{sec:app_net}.

\subsection{Order parameter and critical coupling strength}
\label{sec:critical}

We quantify synchronization of the oscillators using the global  order
parameter~\cite{kuramoto_chemical_1984, acebron_kuramoto_2005, arenas2008synchronization} given by
\begin{equation}
    \label{eq:def_op}
    R(t)
    \coloneqq
    \left| \frac{1}{N}\sum_{j=1}^{N} e^{i\theta_j(t)} \right|.
\end{equation}
The global order parameter satisfies $0\le R(t)\le1$:
$R(t)\simeq1$ indicates nearly complete synchronization, whereas
$R(t)\simeq0$ indicates incoherent oscillator
phases. Because the dynamical system is assumed to be stochastic, we use the temporal average of $R(t)$, denoted by $\langle R\rangle_t$, to reduce the effects of temporal fluctuations in $R(t)$. We provide in Sec.~\ref{sec:numerics} procedures of the temporal averaging for $R(t)$ and other quantities introduced in the following text.

In contrast to the sharp transition in the thermodynamic limit
($N\to\infty$), the synchronization transition in a finite system is rounded,
and $R(t)$ can be nonnegligibly positive below the synchronization transition point~\cite{hong2015finite, rodrigues_kuramoto_2016}. We therefore define the threshold coupling strength for synchronization, $K_\mathrm{c}$, as the smallest value of $K$ that satisfies
\begin{equation}
    \label{eq:def_kcp}
    \frac{\langle R\rangle_t - R_0}{1 - R_0} > p ,
\end{equation}
where $R_0$ denotes $\langle R\rangle_t$ at the first value of $K$ in the
simulation grid (i.e., $K_0 =0.01$). In other words, $K_\mathrm{c}$ is the smallest value of
$K$ at which $\langle R\rangle_t$ has traversed a fraction $p$ of the interval
from its baseline value $R_0$ to complete synchronization
(i.e., $\langle R\rangle_t=1$). Unless otherwise stated, we set $p=0.2$. We confirm in Appendix~\ref{sec:app_change_Kc} that our main conclusions are robust with respect to the choice of $p$ (i.e., $p=0.1$ and $p=0.3$).

Table~\ref{tab:kc} reports the mean and standard deviation of the resulting
$K_\mathrm{c}$ values across ten frequency assignments for each network.
By frequency assignment, we mean to draw of the
natural frequencies $\{\omega_i\}_{i=1}^N$ from $g(\omega)$, independently for the different nodes.
We provide further details of the numerical methods in Sec.~\ref{sec:numerics} and
Appendix~\ref{sec:app_num}.

\begin{table}
    \centering
    \caption{Structural properties and critical coupling strengths of the
    three model networks and six empirical
    networks. The columns give the number of nodes $N$, number of edges $|E|$,
    mean degree $\langle k\rangle$, default size of sentinel sets
    $\lfloor\ln N\rfloor$, and critical coupling
    strength $K_\mathrm{c}$ defined by
    Eq.~\eqref{eq:def_kcp}. Each $K_\mathrm{c}$ entry is the mean over ten independent
    frequency assignments, with the standard deviation in parentheses.}
    \label{tab:kc}
\begin{ruledtabular}
\begin{tabular}{lrrrcr}
Network & \multicolumn{1}{c}{$N$} & \multicolumn{1}{c}{$|E|$} & \multicolumn{1}{c}{$\langle k\rangle$} & \multicolumn{1}{c}{$\lfloor\ln N\rfloor$} & \multicolumn{1}{c}{$K_\mathrm{c}$} \\
\colrule
complete       &  100 &   4950 &  99.00 & 4 & 1.61\,(0.01) \\
ER             &  100 &    514 &  10.28 & 4 & 1.57\,(0.08) \\
scale-free     &  100 &    325 &   6.50 & 4 & 1.44\,(0.18) \\
\colrule
Bat            &   43 &    546 &  25.40 & 3 & 1.26\,(0.13) \\
Highschool boy &   70 &    274 &   7.83 & 4 & 1.45\,(0.18) \\
House finch    &  108 &   1026 &  19.00 & 4 & 1.16\,(0.11) \\
Surfer         &   43 &    336 &  15.63 & 3 & 1.35\,(0.12) \\
Hall           &  217 &   1839 &  16.95 & 5 & 1.47\,(0.12) \\
Jazz           &  198 &   2742 &  27.70 & 5 & 1.23\,(0.11) \\
\end{tabular}
\end{ruledtabular}
\end{table}

\subsection{Early warning signals}
\label{sec:ews}

We calculate the following three types of EWSs.
Because our focus is on EWSs under partial observation, we define these
indices for a sentinel set $S$ of size $n$. We summarize the calculation of each EWS in Figs.~\ref{fig:scheme}(c) and (d). 

First, we use the local order parameter, a
quantity used for analyzing synchronization within a community~\cite{skardal_hierarchical_2012}.
It is defined by
    \begin{equation}
        \label{eq:def_local}
        R_S(t) 
        \coloneqq
        \left| \frac{1}{n}\sum_{j\in S} e^{i\theta_j(t)} \right|
    \end{equation}  
and measures synchrony among the sentinel nodes. We use the temporal average of $R_S(t)$, denoted by $\langle R_S \rangle_t$, as the EWS [see Fig.~\ref{fig:scheme}(c)]. 

Second, we compute the unbiased temporal variance of $R_S(t)$. Its full-observation counterpart (i.e., the temporal variance of the
global order parameter) was previously used as an
EWS~\cite{garcia2017enhancement}.
We denote it by $\mathrm{Var}(R_S)$ [see Fig.~\ref{fig:scheme}(c)].

Third, as an EWS that we propose, we compute the temporal variance of detrended phases. 
To compute this EWS, we start by calculating the residual
phase by subtracting the fitted linear trend from each observed phase:
\begin{equation}
    \label{eq:micro_pre}
    \delta\theta_i(t)
    \coloneqq \theta_i(t)-\hat\omega_i t - \hat\theta_{i,0}. 
\end{equation}
See Figs.~\ref{fig:scheme}(b) and (d).
For this calculation, we treat each phase variable $\theta_i(t)$ as an
unwrapped variable, as shown by the solid curves in Fig.~\ref{fig:scheme}(b), rather than a circular variable confined to $[0,2\pi)$.
We estimate $\hat\omega_i$ and $\hat\theta_{i,0}$ by least-squares fitting of a linear function to the phase time series recorded after the burn-in period [see the dotted lines in Fig.~\ref{fig:scheme}(b)]. We then compute the unbiased temporal variance of $\delta\theta_i(t)$, denoted by $V_i$, for each sentinel node and average it over the sentinel set:
\begin{equation}
    \label{eq:microEWS}
    V_S \coloneqq \frac{1}{n}\sum_{i\in S}\,V_i.
\end{equation}
See Fig.~\ref{fig:scheme}(d).
We use $V_S$ as the third EWS.
Subtracting the linear trend using Eq.~\eqref{eq:micro_pre} removes the mean
rotational motion of each oscillator. Therefore,
$V_i$ measures the amplitude of the residual
phase fluctuations, which is expected
to increase as the onset of synchronization is approached owing to critical slowing down.

\subsection{Selection of sentinel nodes}
\label{sec:sentinel}

To study early warning under partial observations, we calculate the above EWSs using $n$ sentinel nodes. We consider the following six sentinel sets. 
\begin{itemize}[leftmargin=1em]\itemsep1pt
  \item S1: As a full-observation reference, we use all $N$ nodes. The local order parameter in Eq.~\eqref{eq:def_local} is then the global order parameter $R$, and the temporal variance of order parameter [i.e., $\mathrm{Var}(R)$]
  is calculated from $R$. The temporal variance of detrended phases in
  Eq.~\eqref{eq:microEWS} is the average over all nodes.
In contrast to the remaining five sentinel sets, S1 uses all $N$ nodes and is therefore expected to perform better than the remaining sentinel nodes, which use only $n$ ($\ll N$) nodes. We implement S1 as a reference.
  
  \item S2: For each EWS, we rank all nodes using an EWS-specific criterion evaluated at $K=0.9\,K_\mathrm{c}$ and select the top $n$ nodes. For the temporal average and variance of the local order parameter, i.e., $\langle R_S \rangle_t$ and $\mathrm{Var}(R_S)$, we rank the nodes by the extent of phase locking to the mean field, $c_i$.
We define $c_i$ by
    \begin{equation}
        \label{eq:def_ci}
        c_i \coloneqq \left| \left\langle e^{i[\Psi(t)-\theta_i(t)]}\right\rangle_t \right|,
    \end{equation}
    where $\Psi(t)$ is the phase of the global mean field, defined by
    \begin{equation}
        \Psi(t) \coloneqq \mathrm{arg}\, \left[\frac{1}{N}\sum_{j=1}^{N} e^{i\theta_j(t)} \right].
    \end{equation}
Motivated by the correlation index between oscillators~\cite{arenas2006synchronization}, $c_i$ measures the degree of phase locking between node $i$ and the global mean field. The value of $c_i$ is close to $1$ when node $i$ is phase-locked to the mean field, regardless of the phase lag, and close to $0$ when it is little coherent with the mean field. This criterion therefore selects nodes that are genuinely phase-locked to the mean field through network coupling, while it excludes nodes that show temporary alignment with $\Psi(t)$ only because their natural frequencies are close to the mean-field frequency. 

For the temporal variance of detrended phases, $V_S$, we instead rank the nodes by $\mathrm{Var}_t\,\delta\theta_i$. 

  \item S3: We select $n$ nodes using the same EWS-specific criteria as in S2, with the rankings evaluated farther from the onset of synchronization, at
  $K=0.5\,K_\mathrm{c}$.
  \item S4: We select the $n$ nodes with the largest degree
  $k_i\coloneqq\sum_{j=1}^N A_{ij}$ (i.e., hub nodes). This choice follows Ref.~\cite{leyva2025local} and is motivated by the expectation that highly connected nodes may better reflect changes in the collective dynamics. It also allows the sentinel set to be selected solely from the network structure, without requiring dynamical information. 
  \item S5: We select the $n$ nodes with the smallest degree $k_i$.
  \item S6: We use $n$ nodes chosen uniformly at random from all nodes.
\end{itemize}

It should be noted that, unlike S4--S6, constructing S2 and S3 requires observing the time series of all $N$ nodes at one value of $K$.
We only use such $K$ values for assessment purposes;
S2 and S3 represent cases in which this value happens to be close to ($K=0.9K_{\mathrm{c}}$) and
far from ($K=0.5K_{\mathrm{c}}$) the transition, respectively. 
In practice, one can compute the sentinel nodes at an arbitrary value of $K$ whose distance from $K_{\mathrm{c}}$ is unknown.
The following results with S2 and S3 give a guide on how well the three types of EWSs perform
depending on the proximity of the employed $K$ value to $K_{\mathrm{c}}$.

\subsection{Kendall's rank correlation coefficient}
\label{sec:kendall_tau}

As a quality measure for EWSs, we 
use the Kendall's rank correlation coefficient
$\tau$, following previous EWS
studies~\cite{dakos2008slowing,chen_practical_2022}.
It quantifies how close to monotonic the relationship between the coupling strength $K$ and
each EWS is.
For $M$ paired observations
$\{(x_i,y_i)\}_{i=1}^{M}$, where $x_i$ is the control parameter and $y_i$ is the
corresponding EWS, we define Kendall's $\tau$ as~\cite{kendall_treatment_1945}
\begin{equation}
\label{eq:tau}
\tau \coloneqq 
\frac{N_{\mathrm{c}}-N_{\mathrm{d}}}
{\sqrt{(N_0-N_x)(N_0-N_y)}}.
\end{equation}
Here, $N_{\mathrm{c}}$ and $N_{\mathrm{d}}$ are the numbers of concordant and
discordant pairs $(i,j)$, satisfying $(x_i-x_j)(y_i-y_j)>0$ and
$(x_i-x_j)(y_i-y_j)<0$, respectively. Furthermore,
$N_0=M(M-1)/2$, and $N_x$ and $N_y$ are the numbers of tied pairs satisfying $x_i=x_j$ and $y_i=y_j$, respectively.

We evaluate the performance of each EWS by calculating Kendall's $\tau$ between
$K$ and the EWS over the pre-transition range $K_0\leq K\leq K_\mathrm{c}$. A positive
value of $\tau$ indicates that the EWS tends to increase with $K$ and may
therefore be useful for anticipating the transition. Because we increase $K$ in
our simulations, a perfectly monotonic
increase in the EWS yields $\tau=1$.

\subsection{Detection of the critical transition by TIPMOC}
\label{sec:tipmoc}

Although Kendall's $\tau$ quantifies whether an EWS increases with $K$, it
cannot distinguish a gradual linear trend from the accelerating nonlinear
increase expected near a transition. To assess whether an EWS alerts such a nonlinear increase, we apply the tipping-point detector TIPMOC~\cite{masuda2026detecting}.
TIPMOC fits the EWS as a function of the control parameter using a power-law
divergence model and compares this model with a linear model using the corrected
Akaike information criterion ($\mathrm{AICc}$)~\cite{burnham2002model}. 
We define $\Delta\mathrm{AICc}$ as the AICc of the power-law model minus that
of the linear model. TIPMOC fits both models to the data within the expanding
window $[K_0,K]$, whose upper endpoint $K$ advances by one sampled value at
each step. TIPMOC declares a transition when $\Delta\mathrm{AICc}\le -10$ for three
consecutive windows. We denote by $K_{\rm det}$ the upper endpoint $K$ of the window at which TIPMOC declares the transition.  For each sentinel
node set, we apply TIPMOC to each of the three EWSs over the range $K_0 \leq K < K_\mathrm{c}$ and
record whether it detects a transition and, if so, the detection point $K_{\rm det}$. 

\subsection{Numerical simulations}
\label{sec:numerics}
We simulate Eq.~\eqref{eq:kura_network} using the Euler--Maruyama scheme with
time step $\Delta t=0.01$ and noise strength $\sigma=0.15$. Each simulation
runs for $T=300$ time units (TUs), and we discard the first $150$ TUs as
burn-in. After burn-in, we record the observed states every $0.1$ TUs  for a further $150$ TUs and use
them to calculate the temporal average and variance of the global order parameter, $\langle R \rangle_t$ and $\mathrm{Var}(R)$, those of the local order parameter, $\langle R_S \rangle_t$ and $\mathrm{Var}(R_S)$, the temporal variance of detrended phases $V_i$, and the measure of phase locking to the global mean field $c_i$ [Eq.~\eqref{eq:def_ci}].

We draw the natural frequencies, $\omega_i$, from the Gaussian distribution with mean
$0$ and variance $1$ as $g(\omega)$. To avoid sampling fluctuations,
we deterministically construct a set of $N$ frequencies
$\{\tilde \omega_j\}_{j=1}^N$ by~\cite{hong2015finite, rodrigues_kuramoto_2016}
\begin{equation}
    \int_{-\infty}^{\tilde \omega_j} g(\omega)\,d\omega = \frac{j-0.5}{N}.
\end{equation}
For each network, we generate ten independent random permutations of the frequency set $\{\tilde{\omega}_j\}_{j=1}^N$, each defining an assignment of the natural frequencies to the nodes, i.e., $\{\omega_i\}_{i=1}^N$. For each frequency assignment, we perform ten simulations with different random seeds. Thus, we perform
$10\times10=100$ simulations per network in total.

We use the coupling strength $K$ as the sole control parameter. For each
network and frequency assignment, we first determine $K_\mathrm{c}$ as described in
Sec.~\ref{sec:critical} and then perform simulations from $K_0 = 0.01$ to $K_{\mathrm{max}} = 2 K_\mathrm{c}$ to calculate the three EWSs. For sentinel sets S2 and S3, we first select the sentinel nodes through running simulations with an increasing sequence of $K$ from $K_0$ to 
$K=0.9K_{\mathrm{c}}$ and $K=0.5K_{\mathrm{c}}$, respectively.
Then, we run simulations with different random-number seeds from $K_0 = 0.01$ to $K_{\mathrm{max}} = 2 K_\mathrm{c}$ to calculate and evaluate
the EWSs. This procedure is to ensure that sentinel selection and EWS evaluation are independent. For each of the three EWSs, we calculate Kendall's $\tau$ and apply TIPMOC using the EWS values over $K_0 \leq K \leq K_\mathrm{c}$. Appendix~\ref{sec:app_num} provides further details of the numerical
simulations.

\section{Results}
\label{sec:results}
\subsection{Early warning signals from sentinel sets}
\label{sec:result_critical}
\label{sec:result_ews}

\begin{figure*}
\centering
\includegraphics[width=1.\linewidth]{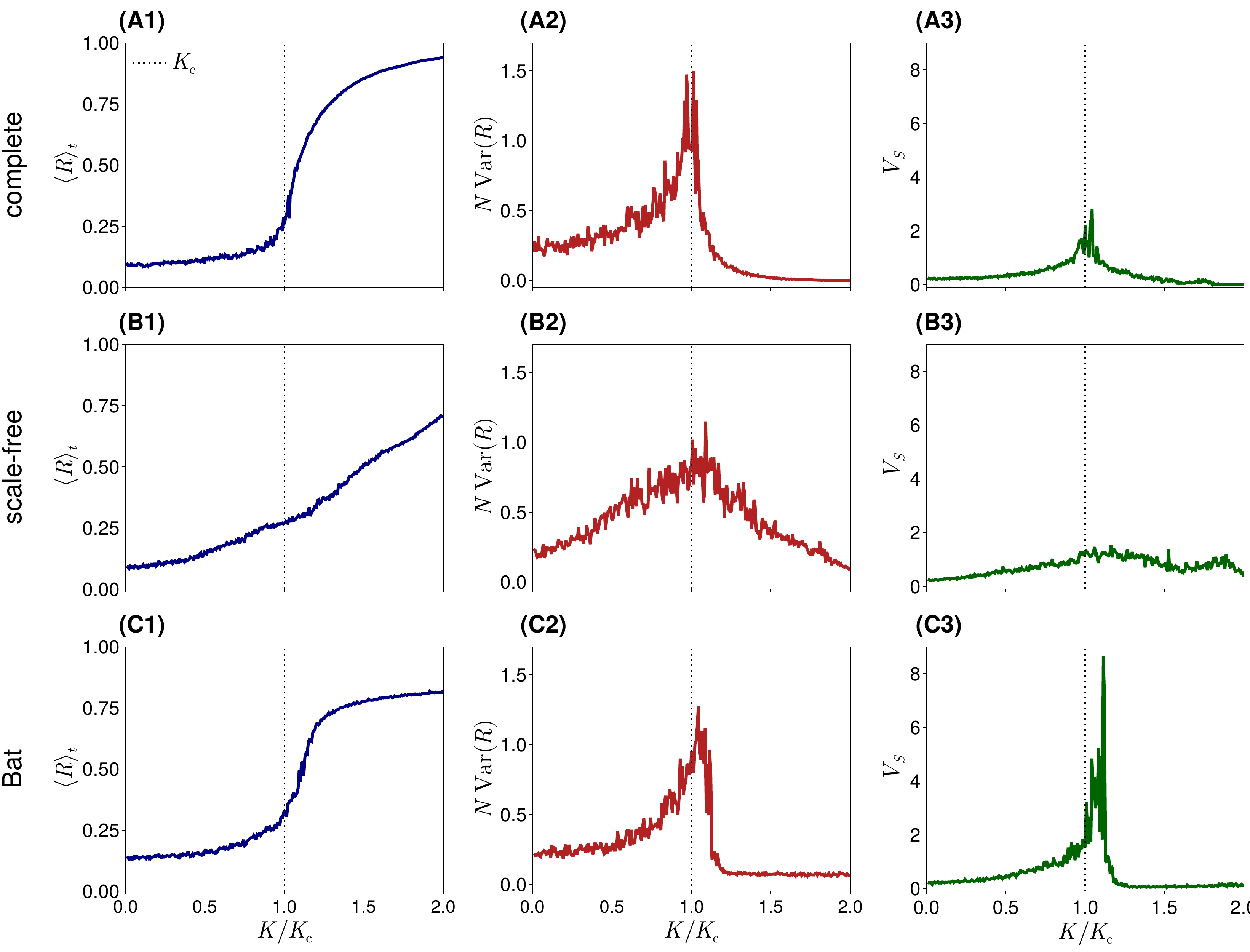}
\caption{
    Three EWSs under full observation (i.e., calculated from all the nodes, i.e., the sentinel set S1) as functions of the rescaled coupling strength $K/K_\mathrm{c}$. Rows show results for the complete graph [panels (A1)--(A3)], a scale-free model network [panels (B1)--(B3)], and the Bat network [panels (C1)--(C3)]. 
    The panels in the left column [panels (A1), (B1), and (C1)] show the time-averaged global order parameter $\langle R\rangle_t$ (i.e., the bifurcation diagram). The panels in the middle column [panels (A2), (B2), and (C2)] show the temporal variance of the scaled global order parameter $N\,\mathrm{Var} (R)$. The panels in the right column [panels (A3), (B3), and (C3)] show the temporal variance of detrended phases $V_S$. 
    Each curve shows a single simulation run. The dotted vertical line marks $K_\mathrm{c}$, i.e., $K/K_\mathrm{c}=1$.}
\label{fig:kc}
\end{figure*}

We first examine the three EWSs under full observation (i.e., sentinel set S1).
Figure~\ref{fig:kc} shows the EWSs as functions of $K/K_\mathrm{c}$ for three
networks. We show the corresponding results for the other six networks in
Fig.~\ref{fig:kc_appendix} in Appendix~\ref{sec:app_remaining}. The left column of
Fig.~\ref{fig:kc} shows the global order parameter $\langle R\rangle_t$, i.e.,
the bifurcation diagrams. The dotted lines show the defined onset of synchronization (i.e., $K = K_\mathrm{c}$), which we aim to anticipate by the EWSs.
The middle and right columns of Fig.~\ref{fig:kc} show the temporal variance
of the global order parameter $\mathrm{Var}(R)$ and the temporal variance of detrended phases $V_S$,
respectively. Both EWSs increase, albeit with fluctuations,
as $K$ approaches $K_\mathrm{c}$ from below and peak around $K_\mathrm{c}$. In the complete graph [Fig.~\ref{fig:kc}(A3)] and the Bat network [Fig.~\ref{fig:kc}(C3)], $V_S$ peaks at coupling strengths slightly above $K_\mathrm{c}$. These offsets arise because $K_\mathrm{c}$ is
defined from the global order parameter $R$ and does not necessarily coincide
with the maxima of the EWS. The other networks show the same qualitative pattern: both $\mathrm{Var}(R)$ and $V_S$ peak around $K_\mathrm{c}$, although the precise peak locations vary depending on the network and the EWS (see Fig.~\ref {fig:kc_appendix}).

\begin{figure*}
    \centering
    \includegraphics[width=1.\textwidth]{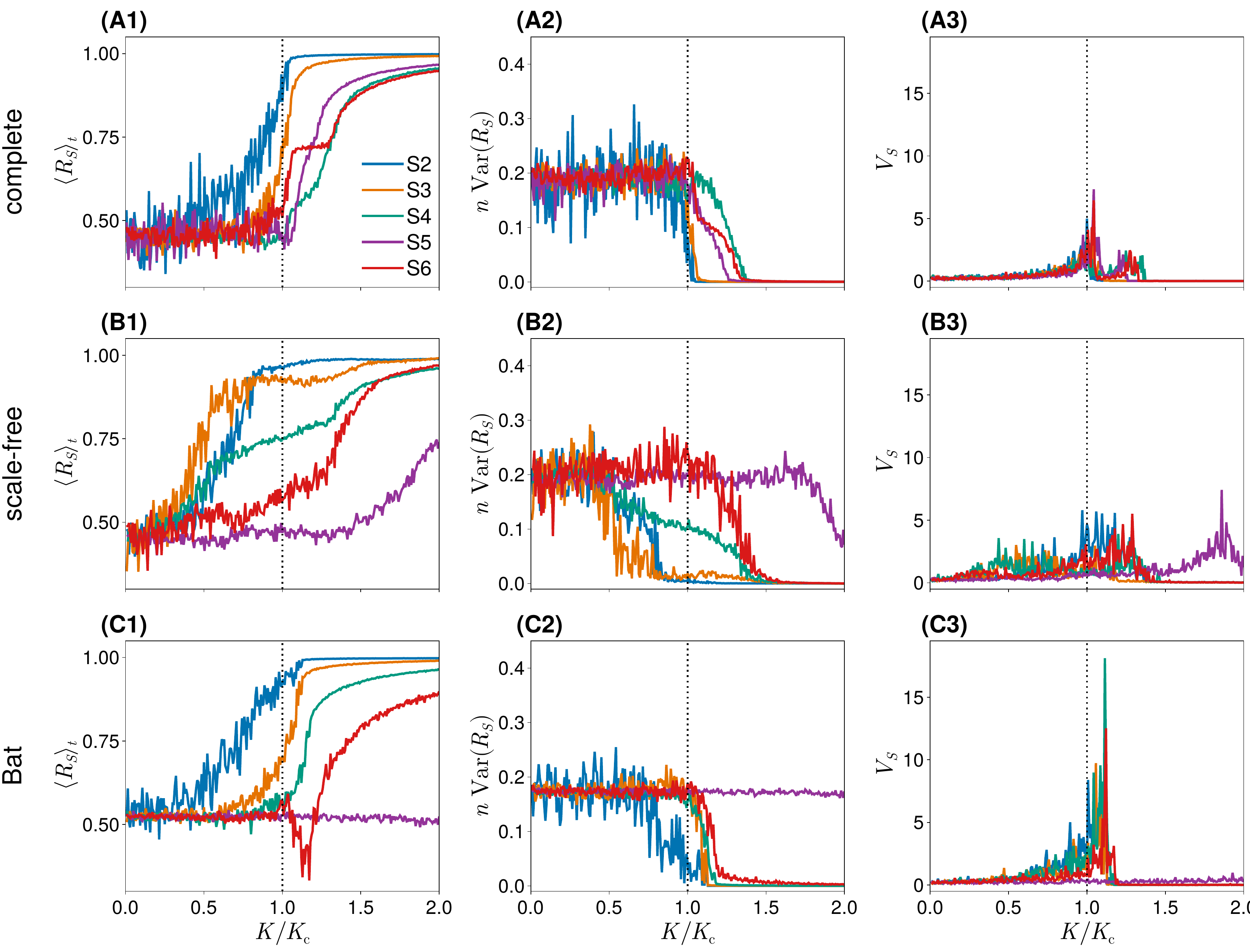}
    \caption{
        Three EWSs under partial observation, computed from sentinel sets
        S2--S6, each containing $n=\lfloor \ln N \rfloor$ nodes. 
        Rows show results for the complete graph [panels (A1)--(A3)], a scale-free model network [panels (B1)--(B3)], and the Bat network [panels (C1)--(C3)]. 
        The panels in the left column [panels (A1), (B1), and (C1)] show the time-averaged local order parameter $\langle R_S\rangle_t$. The panels in the middle column [panels (A2), (B2), and (C2)] show the temporal variance of the scaled local order parameter $n\,\mathrm{Var}(R_S)$. The panels in the right column [panels (A3), (B3), and (C3)] show the temporal variance of detrended phases $V_S$. 
        We omit sentinel set S1 because it uses all nodes, instead of $n$ nodes, and the corresponding results are shown in Fig.~\ref{fig:kc}. Each curve shows a single simulation run. The dotted vertical line marks $K_\mathrm{c}$, i.e., $K/K_\mathrm{c}=1$. 
    }
    \label{fig:ews}
\end{figure*}

Next, we compare the three EWSs under partial observation.
Figure~\ref{fig:ews} shows the EWSs for the same three
networks obtained from sentinel sets S2--S6, each containing
$n=\lfloor\ln N\rfloor$ nodes. We show the corresponding results for
the other six networks in Fig.~\ref{fig:ews_appendix} in
Appendix~\ref{sec:app_remaining}. The qualitative behavior of each EWS is
similar across networks. The local order parameter $\langle R_S\rangle_t$ and
the temporal variance of detrended phases 
$V_S$ increase as $K$ increases, albeit with
fluctuations, over $K_0<K<K_\mathrm{c}$. In contrast, the temporal variance of the local order parameter, $\mathrm{Var}(R_S)$, remains nearly constant or decreases as $K$ increases over
$K_0<K<K_\mathrm{c}$. 

Figures~\ref{fig:ews}(A1), (B1), and (C1) show that $\langle R_S\rangle_t$ begins to
rise at smaller values of $K$ for S2 and S3 than for S5 and S6. This result is
expected because S2 and S3 favor nodes that are relatively strongly
synchronized with the mean field at the selection point, whereas S5 favors low-degree nodes and S6 selects
nodes uniformly at random. The results for S4, which selects hub nodes, depend
on the network structure. For the scale-free network [Fig.~\ref{fig:ews}(B1)], the local order
parameter for S4 rises as rapidly as those for S2 and S3, whereas for the
complete graph and the Bat network [Figs.~\ref{fig:ews}(A1) and (C1)], its curve before $K_\mathrm{c}$ is similar to those for S5 and S6. By contrast, Figs.~\ref{fig:ews}(A2), (B2), and (C2) show that $\mathrm{Var}(R_S)$ remains approximately constant or decreases with $K$ for all sentinel sets. 
Therefore, $\mathrm{Var}(R_S)$ is apparently not good at anticipating synchronization transitions.
Figures~\ref{fig:ews}(A3), (B3), and (C3) show that $V_S$ increases over $K_0<K<K_\mathrm{c}$, while the rising trends differ less among the sentinel sets than those of $\langle R_S\rangle_t$.

\subsection{Evaluating EWSs in terms of Kendall's $\tau$}

To quantify the performance of each EWS and sentinel set, we calculate
Kendall's $\tau$.
Table~\ref{tab:tau_comparison} summarizes the results for the three EWSs across
the nine networks and for two sentinel-set sizes, $n=\lfloor\ln N\rfloor$ and
$n=2\lfloor\ln N\rfloor$. For each combination of EWS and sentinel set, Table~\ref{tab:tau_comparison} reports
the mean and standard deviation over $900$ simulation runs pooled across
the nine networks.

\begin{table*}
    \centering
    \caption{Kendall's $\tau$ for the three EWSs for each sentinel set across the nine networks. Each entry gives the mean of $\tau$
    over $900$ simulations across the nine networks, with the standard
    deviation in parentheses. Because S1 uses all the $N$ nodes, not $n$ nodes, the S1 row is identical between the left and right parts of this table.}
    \label{tab:tau_comparison}
    \setlength{\tabcolsep}{18pt}
    \begin{tabular}{@{\hspace{6pt}}l@{\hspace{6pt}}|ccc|ccc}
    \hline\hline
    \multicolumn{1}{l|}{} & \multicolumn{3}{c|}{$n=\lfloor\ln N\rfloor$} & \multicolumn{3}{c}{$n=2\lfloor\ln N\rfloor$} \\
    \multicolumn{1}{l|}{} & $\langle R_S\rangle_t$ & $\mathrm{Var}(R_S)$ & $V_S$ & $\langle R_S\rangle_t$ & $\mathrm{Var}(R_S)$ & $V_S$ \\
    \hline
    S1 & 0.87\,(0.03) & 0.79\,(0.04) & 0.89\,(0.03) & 0.87\,(0.03) & 0.79\,(0.04) & 0.89\,(0.03) \\
    S2 & 0.76\,(0.08) & $-0.27$\,(0.25) & 0.76\,(0.05) & 0.83\,(0.05) & 0.09\,(0.29) & 0.81\,(0.04) \\
    S3 & 0.65\,(0.17) & $-0.20$\,(0.31) & 0.62\,(0.15) & 0.75\,(0.10) & 0.20\,(0.34) & 0.70\,(0.12) \\
    S4 & 0.44\,(0.37) & $-0.08$\,(0.24) & 0.68\,(0.13) & 0.63\,(0.23) & 0.32\,(0.22) & 0.75\,(0.10) \\
    S5 & 0.05\,(0.17) & 0.01\,(0.10) & 0.34\,(0.25) & 0.12\,(0.22) & 0.07\,(0.14) & 0.47\,(0.26) \\
    S6 & 0.16\,(0.31) & 0.02\,(0.17) & 0.63\,(0.11) & 0.36\,(0.25) & 0.20\,(0.17) & 0.72\,(0.08) \\
    \hline\hline
    \end{tabular}
\end{table*}

The table shows that Kendall's $\tau$ for sentinel set S1
is high for all three EWSs, with mean values ranging from $0.79$ to $0.89$.
Thus, all three EWSs tend to steadily increase toward the synchronization transition under full
observation. Under partial observation (i.e., S2--S6) with $n=\lfloor\ln N\rfloor$ sentinel
nodes (the left part of Table~\ref{tab:tau_comparison}), $\tau$ is smaller than in the case of full observation for all EWSs and all sentinel set selection methods. This result is expected because S2--S6 use only $n$ sentinel nodes, whereas S1 uses all the $N$ nodes.

Under partial observation, $\tau$ with $\mathrm{Var}(R_S)$ is small for any sentinel set selection method, S2--S6.
Between the other two EWSs, which perform consistently better than $\mathrm{Var}(R_S)$ in terms of $\tau$,
the performance is similar for S2. For S3, $\langle R_S \rangle_t$
performs better than $V_S$. The opposite is the case for S4, S5, and S6. Among S2--S6, S2 performs the best
for both $\langle R_S \rangle_t$ and $V_S$, and their average $\tau$ values are 0.11 and 0.13 smaller, respectively than those for S1. When S3--S6 are used, this gap is at least $0.22$ and $0.21$ for 
$\langle R_S \rangle_t$ and $V_S$, respectively.
Considering that S2--S6 use only $n$ sentinel nodes, we regard that S2 provides EWSs of reasonable quality when our ability to observe various nodes is limited.

The right part of Table~\ref{tab:tau_comparison} shows that doubling the sentinel-set size to
$n=2\lfloor\ln N\rfloor$ preserves these qualitative patterns while increasing
$\tau$ for the partial-observation sets S2--S6. We highlight that, with $n=2\lfloor\ln N\rfloor$ sentinel nodes, which is still much fewer than $N$ nodes,
the $\tau$ values for S2 are substantially closer to those under full observation (i.e., S1) if the EWS is $\langle R_S \rangle_t$ or $V_S$.

\subsection{Detection of the transition with TIPMOC}
\label{sec:result_tipmoc}

Finally, we use TIPMOC to assess whether
each EWS detects the impending synchronization transition. We apply TIPMOC to every
combination of the EWS, network, and sentinel set. Table~\ref{tab:tipmoc_detection} reports
the detection rates for sentinel-set sizes $n=\lfloor\ln N\rfloor$ and
$n=2\lfloor\ln N\rfloor$. We calculate the detection rate as the fraction of the $900$ simulations ($9$ networks $\times$ $100$ simulations per network) in which TIPMOC successfully detects the onset of synchronization before $K \in [K_0, K_\mathrm{c})$ reaches $K_\mathrm{c}$. 

Under full observation (i.e., S1), all three EWSs have a detection rate of
$100\%$. Under partial observation with $n=\lfloor\ln N\rfloor$ sentinel nodes,
the left part of Table~\ref{tab:tipmoc_detection} shows that the detection rate
for $\mathrm{Var}(R_S)$ is
at most $5\%$ for every sentinel set. This poor performance is consistent with the results for the Kendall's $\tau$ shown in Table~\ref{tab:tau_comparison}.

For $\langle R_S\rangle_t$, the sentinel set S2 yields the highest detection rate, followed
by S4, which consists of hub nodes. With S3, i.e., when the sentinel set is optimized farther
from the transition than with S2, the detection rate yet remains higher than that for
the purely random sentinel set S6. For 
$V_S$, the ranking across
sentinel sets S2--S6 is similar to that of $\langle R_S\rangle_t$, whereas $V_S$ yields a notably higher detection rate than $\langle R_S\rangle_t$ for every sentinel set. For both $\langle R_S\rangle_t$
and $V_S$, the detection rate with
sentinel set S2 is comparable to that under full observation (i.e., S1).

To examine how early the detection occurs when it does, we show the distribution of the normalized detection point $K_{\rm det}/K_\mathrm{c}$ in Appendix~\ref{sec:app_kdet}. For S2, the median of $K_{\rm det}/K_\mathrm{c}$ among the runs
in which TIPMOC detects the transition before $K_\mathrm{c}$ is 0.70 for $\langle R_S\rangle_t$ and 0.63 for $V_S$.

The right part of Table~\ref{tab:tipmoc_detection} shows that doubling the
sentinel-set size qualitatively preserves these patterns while increasing the
detection rates for S2--S6, similarly to the results for Kendall's $\tau$. Despite using only
$n=2\lfloor\ln N\rfloor$ sentinel nodes, which is still much fewer than the full $N$ nodes, the detection rates for
$\langle R_S\rangle_t$ with S2 and for
$V_S$ with S2--S4 are close to or are equal to 100\%.

As a final robustness test, we have confirmed that $\tau$ and the TIPMOC detection rate are
qualitatively the same when we define $K_\mathrm{c}$ with $p=0.1$ or $p=0.3$ instead of $p=0.2$ (see Appendix~\ref{sec:app_change_Kc}).

In sum, using S2 yields the TIPMOC detection rates
close to those of S1 for both $\langle R_S\rangle_t$ and $V_S$ although full observation (i.e., S1) always beats partial observation (i.e., S2--S6).
In addition, as we mentioned above, $V_S$ exceeds
$\langle R_S\rangle_t$ for every
partial-observation sentinel set S2--S6 in terms of the detection rate.

\begin{table*}
    \centering
    \caption{TIPMOC detection rates (\%) for each combination of sentinel set
    and EWS across the nine networks. Each entry gives the percentage of
    successful detections among $900$ simulations across the nine networks.}
    \label{tab:tipmoc_detection}
    \setlength{\tabcolsep}{26pt}
    \begin{tabular}{@{\hspace{6pt}}l@{\hspace{6pt}}|ccc|ccc}
    \hline\hline
    \multicolumn{1}{l|}{} & \multicolumn{3}{c|}{$n=\lfloor\ln N\rfloor$} & \multicolumn{3}{c}{$n=2\lfloor\ln N\rfloor$} \\
    \multicolumn{1}{l|}{} & $\langle R_S\rangle_t$ & $\mathrm{Var}(R_S)$ & $V_S$ & $\langle R_S\rangle_t$ & $\mathrm{Var}(R_S)$ & $V_S$ \\
    \hline
    S1 & 100 & 100 & 100 & 100 & 100 & 100 \\
    S2 & 90 & 1 & 100 & 98 & 13 & 100 \\
    S3 & 56 & 2 & 88 & 75 & 8 & 98 \\
    S4 & 76 & 5 & 91 & 90 & 35 & 97 \\
    S5 & 9 & 0 & 35 & 19 & 4 & 52 \\
    S6 & 38 & 2 & 82 & 61 & 12 & 94 \\
    \hline\hline
    \end{tabular}
\end{table*}

\section{Discussion}
\label{sec:discussion}

Using the stochastic Kuramoto model on networks,
we numerically investigated EWSs computed from small sets of sentinel nodes, aiming at anticipating synchronization transitions. We introduced the temporal variance of
detrended phases [$V_S$; Eq.~\eqref{eq:microEWS}] as an EWS and compared it with two
previously used EWSs, $\langle R_S \rangle_t$ and $\mathrm{Var}(R_S)$. We found that $\langle R_S \rangle_t$ and $V_S$
were consistently better than $\mathrm{Var}(R_S)$.
Furthermore, the proposed $V_S$ performed overall better than
$\langle R_S \rangle_t$, in terms of the rate to detect synchronization transitions, while they showed
comparably high performance in terms of Kendall's $\tau$.
Moreover, for both $\langle R_S \rangle_t$ and $V_S$, sentinel set S2 consistently
outperformed the other sentinel set heuristics (S3--S6), including the
degree-based S4 strategy motivated by Ref.~\cite{leyva2025local}.
Thus, when node dynamics are heterogeneous,
we suggest that sentinel selection uses the information on heterogeneity in the dynamics of nodes (such as in S2, or its harsh variant S3), potentially in combination with
network structure if available. Because constructing S2 or S3 requires observation of all nodes at one value of $K$, developing practical
selection criteria from limited observations remains an important direction
for future work.

Previous studies of EWS detection for the Kuramoto model under partial observation remain limited. Liu et al. developed a supervised-learning approach for anticipating systemic transitions in networked dynamics, although partial observation was not specifically examined for the Kuramoto model~\cite{liu2024early}. Unlike their approach, which requires extensive training data and targets full phase locking, our method requires relatively little data and thus may be useful in low-data situations. 

One direction for extending the present work is to anticipate diverse dynamics in coupled oscillatory systems, including cluster synchronization~\cite{pecora_cluster_2014}, chimera states~\cite{kuramoto_coexistence_2002,abrams_chimera_2004}, and spiral waves~\cite{shima_rotating_2004,ottino-loffler_frequency_2016}. Previous studies have predicted chimera states and cluster synchronization using reservoir computing~\cite{chauhan_predicting_2025}, identified the precursor of the collapse of chimera states~\cite{andrzejak_all_2016}, and identified indicators for detecting transitions from stable spiral waves to spatiotemporal chaos~\cite{seenivasan_using_2016}. A relevant open question is whether the EWSs and sentinel-selection methods developed in this study can anticipate such diverse transitions based on observations from only a few nodes.

Another direction of future research is to evaluate whether the EWSs considered in this study can anticipate synchronization transitions in experimental data. A publicly available dataset of coupled nonlinear electronic oscillators provides a possible benchmark because it contains measurements across multiple network structures and coupling strengths~\cite{vera-avila_experimental_2020}. A biologically relevant example may be single-cell recordings from the mouse suprachiasmatic nucleus, in which the expressions of clock genes are measured in individual cells~\cite{yamaguchi_synchronization_2003,abel_functional_2016}. In both cases, the recorded variable is a periodic observable rather than the oscillator phase; therefore, the phase [i.e., $\theta_i$ in Eq.~\eqref{eq:kura_network}] must first be inferred from the observed time series~\cite{matsuki_extended_2023}. Another possible application is anticipating loss of synchrony in power grids~\cite{witthaut2022collective}. Unlike electronic-oscillator and circadian data, power-grid phase angles can be obtained directly from phasor measurement units, and large volumes of such measurement data are publicly available~\cite{biswas_open-source_2023}. 

\begin{acknowledgments}
N.M.~acknowledges support from the Japan Science and Technology Agency (JST) Moonshot R\&D Grant Number JPMJMS2021, the National Science Foundation (under grant no.~2204936), and JSPS KAKENHI (under grant nos.~JP 23H03414, 24K14840, and 24K03013).

The authors used ChatGPT 5.6 Sol and Claude Opus 4.8 to assist with reviewing and developing the code, and revising the  manuscript. All scientific content, interpretations, code, and manuscript text were reviewed and verified by the authors. 
\end{acknowledgments}

\section{Data availability}
The data that support the findings of this article are openly available~\cite{our_code}.

\appendix

\section{Networks}
\label{sec:app_net}

\subsection{Model networks}

We generate all model networks with $N=100$ nodes as follows.
\begin{description}
    \item[Complete graph] In the complete graph, every pair of
    nodes is connected.
    \item[Erd\H{o}s--R\'enyi random graph] We construct an Erd\H{o}s--R\'enyi
    random graph~\cite{erdos1959random} with edge probability $p=10/(N-1)$, so
    that the mean degree is $10$. We repeat
    the network generation until the resulting
    network is connected.
    \item[Scale-free network] We generate a scale-free network with degree
    exponent $\gamma=3$ using the configuration model~\cite{newman_networks_2018}.
    The degree of each node is drawn independently from the truncated power-law
    distribution $p(k)\propto k^{-\gamma}$ with $k_{\min}\le k\le N-1$. We set
    $k_{\min}=4$ to prevent isolated nodes from appearing with high probability. If the total degree is odd, we increase the degree of a minimum-degree node by one. We construct the network from
    this power-law degree sequence using a Julia
    package~\cite{graphjl}. We repeat the network generation until the
    resulting network is connected.
\end{description}

\subsection{Empirical networks}

We take all empirical networks from the \texttt{networkdata} R
package~\cite{schoch2022networkdata} using the dataset names specified below
(e.g., ``animal\_18''). We convert each network into a simple, undirected, and
unweighted network and use only its largest connected component. The short
network names (e.g., ``Bat'') follow Ref.~\cite{maclaren2023early}.
\begin{description}
    \item[Bat] The first network in the ``animal\_18'' dataset~\cite{silvis_roosting_2014}, representing associations
    among bats ({\it Myotis sodalis}). This network has $43$ nodes and $546$
    edges.
    \item[Highschool boy] The ``highschool\_boys'' network~\cite{Coleman_introduction_1964}, representing
    friendship ties among high-school students. This network has $70$
    nodes and $274$ edges.
    \item[House finch] The ``animal\_6'' network~\cite{adelman_feeder_2015}, representing associations
    among house finches ({\it Haemorhous mexicanus}). This network has $108$
    nodes and $1026$ edges.
    \item[Surfer] The ``surfersb'' network~\cite{freeman_human_1988}, representing social interactions in
    a windsurfing community. This network has $43$ nodes and $336$ edges.
    \item[Hall] The ``hall'' network~\cite{freeman_exploring_1998}, representing friendship interactions among
    residents of a residence hall. This network has $217$ nodes and $1839$
    edges.
    \item[Jazz] The ``jazz'' network~\cite{gleiser_community_2003}, representing collaborations among jazz
    bands. This network has $198$ nodes and $2742$ edges.
\end{description}

\section{Details of numerical simulations}
\label{sec:app_num}

For each network and frequency assignment, we first determine $K_\mathrm{c}$ from
chained up-sweeps over $K_0=0.01\leq K\leq 3.0$. Starting from $K=K_0$, we
increase $K$ in increments of $0.01$. At $K_0$, we draw the initial phase of
each oscillator independently and uniformly from $[0,2\pi)$. At each
subsequent value of $K$, we use the final state at the preceding value as the
initial condition. We perform $10$ independent up-sweeps with different random
seeds. At each $K$, we average $\langle R\rangle_t$ across the ten runs to
obtain the run-average, denoted by $\langle \langle R\rangle_t \rangle_{\mathrm{run}}$. We then
apply the threshold given by Eq.~\eqref{eq:def_kcp} to $\langle \langle R\rangle_t \rangle_{\mathrm{run}}$ to determine $K_\mathrm{c}$.

For the EWS analyses, we perform ten independent chained up-sweeps over the range $K_0\leq K\leq 2K_\mathrm{c}$ for each network and frequency assignment. We carry out the sentinel-node selection described in Sec.~\ref{sec:sentinel} independently for every combination of network, frequency assignment, and chained up-sweep. For S2 and S3, which require observing all nodes at one value of $K$, we use chained up-sweeps with random seeds different from those used for simulating and evaluating the EWSs. Therefore, the sentinel selection and the evaluation of the EWSs are based on independent realizations of the initial conditions and dynamical noise. For the sentinel sets S2--S5, ties in the node rankings are broken at random.

\section{Results for the other six networks}

\label{sec:app_remaining}
\begin{figure*}
\centering
\includegraphics[width=.95\linewidth]{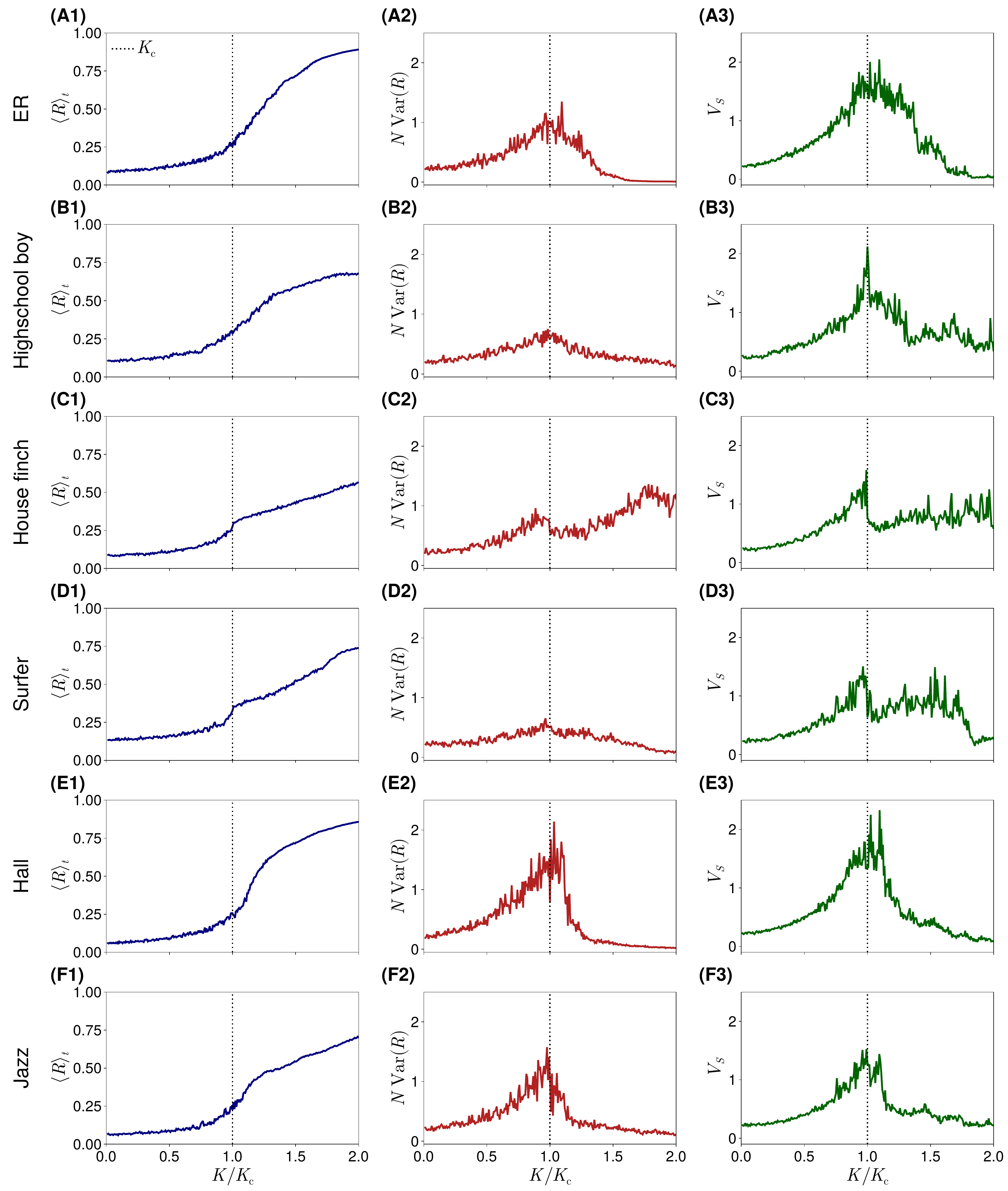}
\caption{
    Three EWSs under full observation for the six networks not shown
    in Fig.~\ref{fig:kc}. Rows show results for an ER model network [panels (A1)--(A3)],
    the Highschool boy network [panels (B1)--(B3)], the House finch network [panels (C1)--(C3)], the Surfer network [panels (D1)--(D3)], the Hall network [panels (E1)--(E3)], and the Jazz network [panels (F1)--(F3)]. 
    The panels in the left column show the time-averaged global order parameter $\langle R\rangle_t$ (i.e., the bifurcation diagram). The panels in the middle column show the temporal variance of the scaled global order parameter $N\,\mathrm{Var}(R)$. The panels in the right column show the temporal variance of detrended phases $V_S$. Each curve shows a single simulation run.
    The dotted vertical line marks $K_\mathrm{c}$, i.e., $K/K_\mathrm{c}=1$.
    }
\label{fig:kc_appendix}
\end{figure*}

\begin{figure*}
\centering
\includegraphics[width=.95\textwidth]{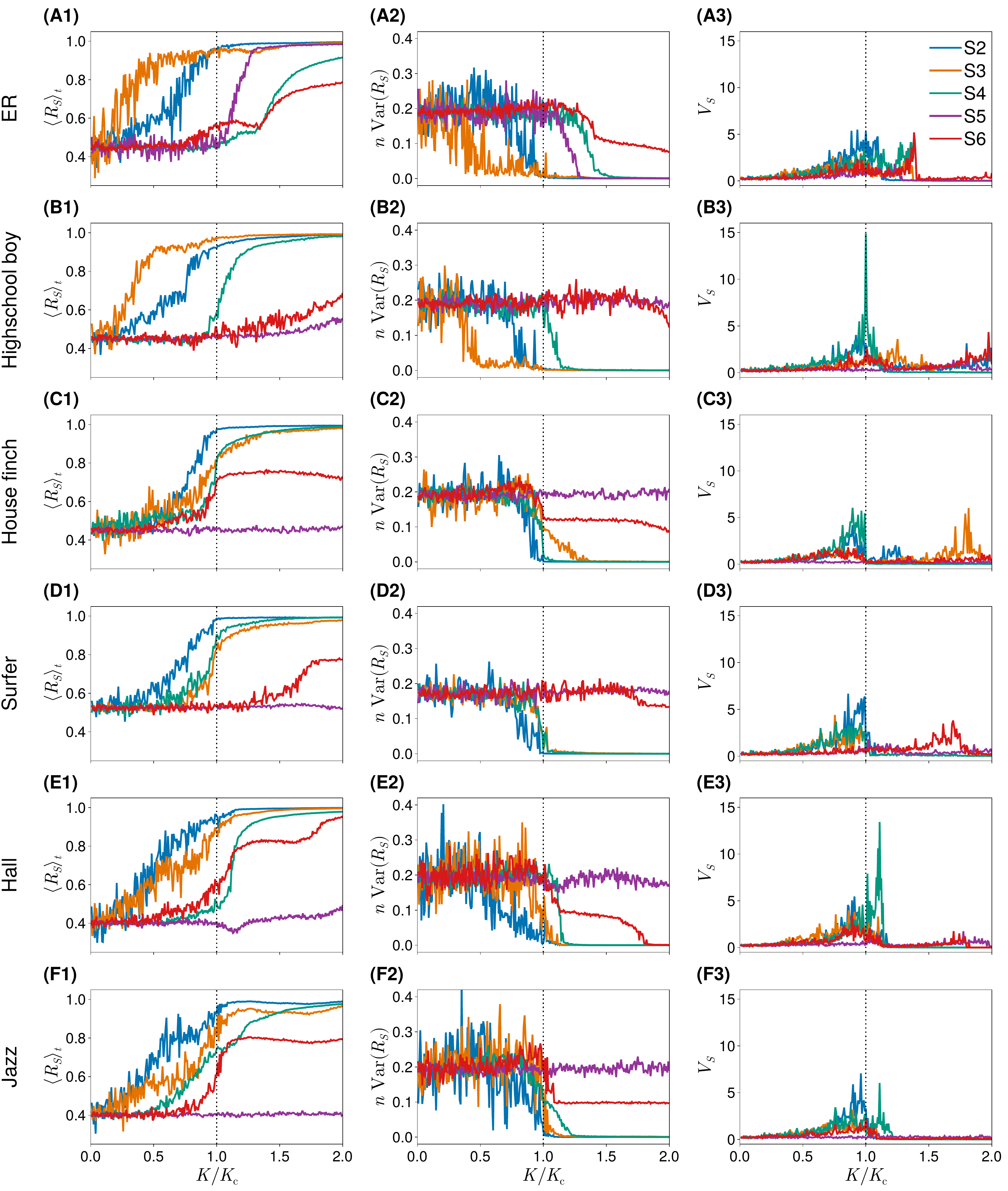}
\caption{ Three EWSs under partial observation
for the six networks not shown in Fig.~\ref{fig:ews}.
We computed the EWSs from sentinel sets S2--S6, each containing $n=\lfloor \ln N \rfloor$ nodes. Rows show results for an ER model network [panels (A1)--(A3)], the Highschool boy network [panels (B1)--(B3)], the House finch network [panels (C1)--(C3)], the Surfer network [panels (D1)--(D3)], the Hall network [panels (E1)--(E3)], and the Jazz network [panels (F1)--(F3)]. 
The panels in the left column show the time-averaged local order parameter $\langle R_S\rangle_t$. The panels in the middle column show the temporal variance of the scaled local order parameter $n\,\mathrm{Var}(R_S)$. The panels in the right column show the temporal variance of detrended phases $V_S$. Each curve shows a single simulation run. The dotted vertical line marks $K_\mathrm{c}$, i.e., $K/K_\mathrm{c}=1$. 
}
\label{fig:ews_appendix}
\end{figure*}

Figures~\ref{fig:kc} and \ref{fig:ews} in the main text show
the results for the complete graph, scale-free network, and Bat network.
Figures~\ref{fig:kc_appendix} and \ref{fig:ews_appendix} present the corresponding results for the remaining
six networks. The results are similar to those for the three networks shown in Figs.~\ref{fig:kc} and \ref{fig:ews}.

\section{Detection points of TIPMOC}
\label{sec:app_kdet}

\begin{figure*}
  \includegraphics[width=.95\linewidth]{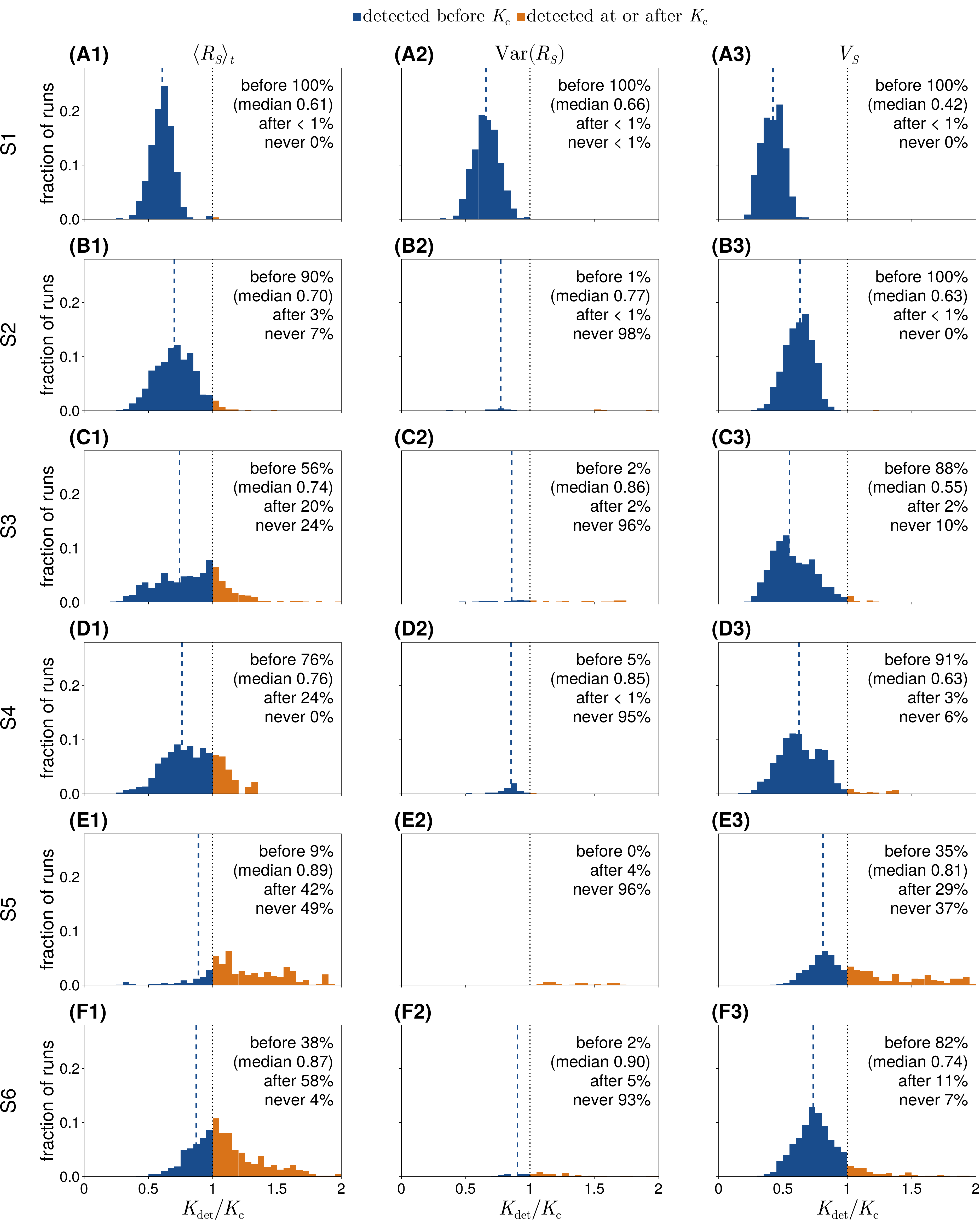}
    \caption{Histograms of the TIPMOC detection point $K_{\rm det}/K_\mathrm{c}$.
    For each sentinel set with $n=\lfloor \ln N \rfloor$, we apply TIPMOC to each of the three EWSs over the range $K_0 \le K \le 2K_\mathrm{c}$.
    Rows correspond to sentinel sets S1--S6. The panels in the left, middle, and right columns show the results
    for $\langle R_S\rangle_t$, $\mathrm{Var}(R_S)$, and $V_S$, respectively.
    The dotted vertical line marks $K_{\rm det}=K_\mathrm{c}$.
    Blue and orange bars correspond to the runs with $K_{\rm det} < K_\mathrm{c}$ and $K_{\rm det} \ge K_\mathrm{c}$,
    respectively. The vertical axis shows the fraction of the 900 runs across the nine networks.
    The numbers in each panel give the percentages of runs with $K_{\rm det} < K_\mathrm{c}$ (``before''),
    $K_{\rm det} \ge K_\mathrm{c}$ (``after''), and no detection up to $K=2K_\mathrm{c}$ (``never'').
    The dashed vertical line and the number in parentheses indicate the median of $K_{\rm det}/K_\mathrm{c}$
    over the runs with $K_{\rm det} < K_\mathrm{c}$ (i.e., the blue bars).
    No median is shown when no run detects the transition before $K_\mathrm{c}$ [panel (E2)].
    }
  \label{fig:kdet}
\end{figure*}

Figure~\ref{fig:kdet} shows the distribution of the TIPMOC detection point normalized by $K_\mathrm{c}$,
i.e., $K_{\rm det}/K_\mathrm{c}$, for $n=\lfloor \ln N \rfloor$.
In this analysis, we continue to apply TIPMOC beyond $K_\mathrm{c}$ up to $K=2K_\mathrm{c}$ to distinguish detections at or after $K_\mathrm{c}$ from no detection.
Under full observation (i.e., S1), all three EWSs detect the transition before $K_\mathrm{c}$ in almost all runs [Figs.~\ref{fig:kdet}(A1), (A2), and (A3)]. 
For S2, TIPMOC tends to detect the transition later than for S1 with $\langle R_S\rangle_t$ and $V_S$ [Figs.~\ref{fig:kdet}(B1) and (B3)].
However, the detections still typically occur well before $K_\mathrm{c}$, with median $K_{\rm det}/K_\mathrm{c}$ values of 0.70 and 0.63, respectively, among the runs with detection before $K_\mathrm{c}$.

We cannot evaluate false alarms, i.e., alarms in the absence of an impending transition, since we increase $K$ toward the synchronization transition in all simulations.
Nevertheless, our results suggest that TIPMOC rarely raises alarms when the EWS does not increase with $K$.
Indeed, for $\mathrm{Var}(R_S)$ with S2--S6, Kendall's $\tau$ is close to zero (Table~\ref{tab:tau_comparison}),
and TIPMOC detects a transition before $K_\mathrm{c}$ in at most 5\% of the runs [Figs.~\ref{fig:kdet}(B2), (C2), (D2), (E2), and (F2)].

\section{Different thresholds for defining $K_\mathrm{c}$}
\label{sec:app_change_Kc}

To examine whether our results depend on the threshold $p$ used for defining the critical coupling strength $K_\mathrm{c}$ [Eq.~\eqref{eq:def_kcp}],
we repeat the analysis with $p=0.1$ and $p=0.3$.
For each value of $p$, we recompute $K_\mathrm{c}$, reselect the sentinel nodes, perform $100$ simulations for each of the nine networks, and compute the three EWSs for the six sentinel sets with $n=\lfloor \ln N \rfloor$ and $n=2\lfloor \ln N \rfloor$, as in the case of $p=0.2$.

Table~\ref{tab:kc_dif_p} lists the values of $K_\mathrm{c}$ for $p=0.1$, $0.2$, and $0.3$. As expected, $K_\mathrm{c}$ increases with $p$ for all networks. Table~\ref{tab:tau_comparison_p} shows Kendall's $\tau$ for $p=0.1$ and $p=0.3$ and Table~\ref{tab:tipmoc_detection_p} shows the corresponding TIPMOC detection rates.
For both values of $p$ and both sentinel-set sizes, $\langle R_S\rangle_t$ and $V_S$ outperform $\mathrm{Var}(R_S)$,
and S2 yields the best performance among S2--S6, as in the case of $p=0.2$.

\begin{table}[h]
    \centering
    \caption{Critical coupling strength, $K_\mathrm{c}$, for $p=0.1$, $0.2$, and $0.3$. Each entry is the mean over ten independent frequency assignments, with the standard deviation in parentheses.}
    \label{tab:kc_dif_p}
    \begin{ruledtabular}
    \begin{tabular}{lrrr}
    Network & \multicolumn{1}{c}{$p=0.1$} & \multicolumn{1}{c}{$p=0.2$} & \multicolumn{1}{c}{$p=0.3$} \\
    \colrule
    complete       &  1.45\,(0.02) & 1.61\,(0.01) & 1.67\,(0.00) \\
    ER             &  1.24\,(0.08) & 1.57\,(0.08) & 1.75\,(0.10) \\
    scale-free     &  1.05\,(0.14) & 1.44\,(0.18) & 1.67\,(0.16) \\
    \colrule
    Bat            &  1.10\,(0.12) & 1.26\,(0.13) & 1.37\,(0.15) \\
    Highschool boy &  1.05\,(0.11) & 1.45\,(0.18) & 1.66\,(0.22) \\
    House finch    &  0.97\,(0.14) & 1.16\,(0.11) & 1.35\,(0.06) \\
    Surfer         &  1.10\,(0.07) & 1.35\,(0.12) & 1.49\,(0.16) \\
    Hall           &  1.20\,(0.09) & 1.47\,(0.12) & 1.64\,(0.13) \\
    Jazz           &  1.06\,(0.08) & 1.23\,(0.11) & 1.39\,(0.08) \\
    \end{tabular}
    \end{ruledtabular}
\end{table}

\begin{table*}
    \centering
    \caption{Kendall's $\tau$ for the three EWSs for each sentinel set when $K_\mathrm{c}$ is defined with $p=0.1$ and $p=0.3$. Each entry gives the mean of $\tau$ over $900$ simulations across the nine networks, with the standard deviation in parentheses.}
    \label{tab:tau_comparison_p}
    \setlength{\tabcolsep}{14pt}
    \begin{tabular}{@{\hspace{4pt}}l@{\hspace{10pt}}@{\hspace{6pt}}l@{\hspace{10pt}}|ccc|ccc}
    \hline\hline
    \multicolumn{1}{l}{} & & \multicolumn{3}{c|}{$n=\lfloor\ln N\rfloor$} & \multicolumn{3}{c}{$n=2\lfloor\ln N\rfloor$} \\
    \multicolumn{1}{l}{} & & $\langle R_S\rangle_t$ & $\mathrm{Var}(R_S)$ & $V_S$ & $\langle R_S\rangle_t$ & $\mathrm{Var}(R_S)$ & $V_S$ \\
    \hline
    \multirow{6}{*}{$p = 0.1$}
    & S1 & 0.83\,(0.03) & 0.75\,(0.05) & 0.88\,(0.03) & 0.83\,(0.03) & 0.75\,(0.05) & 0.88\,(0.03) \\
    & S2 & 0.68\,(0.13) & $-0.12$\,(0.26) & 0.74\,(0.05) & 0.77\,(0.07) & 0.29\,(0.26) & 0.79\,(0.04) \\
    & S3 & 0.53\,(0.20) & $-0.02$\,(0.27) & 0.66\,(0.10) & 0.65\,(0.12) & 0.35\,(0.20) & 0.73\,(0.07) \\
    & S4 & 0.32\,(0.35) & 0.03\,(0.19) & 0.69\,(0.09) & 0.52\,(0.25) & 0.32\,(0.19) & 0.76\,(0.06) \\
    & S5 & 0.03\,(0.14) & 0.01\,(0.10) & 0.28\,(0.22) & 0.08\,(0.17) & 0.05\,(0.11) & 0.40\,(0.24) \\
    & S6 & 0.09\,(0.24) & 0.03\,(0.13) & 0.58\,(0.11) & 0.23\,(0.23) & 0.13\,(0.16) & 0.68\,(0.07) \\
    \hline
    \multirow{6}{*}{$p = 0.3$}
    & S1 & 0.89\,(0.03) & 0.78\,(0.06) & 0.87\,(0.05) & 0.89\,(0.03) & 0.78\,(0.06) & 0.87\,(0.05) \\
    & S2 & 0.78\,(0.09) & $-0.36$\,(0.22) & 0.75\,(0.06) & 0.85\,(0.05) & $-0.05$\,(0.28) & 0.80\,(0.04) \\
    & S3 & 0.71\,(0.16) & $-0.32$\,(0.31) & 0.51\,(0.22) & 0.78\,(0.11) & 0.02\,(0.37) & 0.61\,(0.19) \\
    & S4 & 0.52\,(0.36) & $-0.21$\,(0.25) & 0.60\,(0.18) & 0.69\,(0.22) & 0.19\,(0.27) & 0.68\,(0.17) \\
    & S5 & 0.06\,(0.21) & $-0.00$\,(0.10) & 0.36\,(0.25) & 0.14\,(0.25) & 0.07\,(0.15) & 0.49\,(0.26) \\
    & S6 & 0.22\,(0.34) & $-0.03$\,(0.19) & 0.61\,(0.14) & 0.44\,(0.24) & 0.21\,(0.18) & 0.71\,(0.10) \\
    \hline\hline
    \end{tabular}
\end{table*}

\begin{table*}
    \centering
    \caption{TIPMOC detection rates (\%) for each combination of sentinel set
    and EWS when $K_\mathrm{c}$ is defined with $p=0.1$ and $p=0.3$. Each entry gives the percentage of
    successful detections among $900$ simulations across the nine networks.}
    \label{tab:tipmoc_detection_p}
    \setlength{\tabcolsep}{24pt}
    \begin{tabular}{@{\hspace{4pt}}l@{\hspace{10pt}}@{\hspace{6pt}}l@{\hspace{10pt}}|ccc|ccc}
    \hline\hline
    \multicolumn{1}{l}{} & & \multicolumn{3}{c|}{$n=\lfloor\ln N\rfloor$} & \multicolumn{3}{c}{$n=2\lfloor\ln N\rfloor$} \\
    \multicolumn{1}{l}{} & & $\langle R_S\rangle_t$ & $\mathrm{Var}(R_S)$ & $V_S$ & $\langle R_S\rangle_t$ & $\mathrm{Var}(R_S)$ & $V_S$ \\
    \hline
    \multirow{6}{*}{$p = 0.1$}
    & S1 & 99 & 96 & 100 & 99 & 96 & 100 \\
    & S2 & 70 & 1 & 96 & 89 & 8 & 100 \\
    & S3 & 30 & 0 & 70 & 44 & 3 & 86 \\
    & S4 & 49 & 3 & 83 & 63 & 17 & 95 \\
    & S5 & 3 & 0 & 19 & 5 & 0 & 31 \\
    & S6 & 12 & 0 & 61 & 18 & 1 & 82 \\
    \hline
    \multirow{6}{*}{$p = 0.3$}
    & S1 & 100 & 100 & 100 & 100 & 100 & 100 \\
    & S2 & 94 & 2 & 100 & 99 & 17 & 100 \\
    & S3 & 72 & 2 & 94 & 86 & 11 & 99 \\
    & S4 & 87 & 5 & 92 & 95 & 38 & 97 \\
    & S5 & 19 & 0 & 42 & 30 & 8 & 59 \\
    & S6 & 54 & 4 & 88 & 80 & 21 & 96 \\
    \hline\hline
    \end{tabular}
\end{table*}

\bibliography{Laplacian_EWS}

\end{document}